\documentclass[preprint,12pt]{elsarticle}

\usepackage{amssymb}
\usepackage{amsmath}
\usepackage{multirow}
\usepackage{xurl}
\usepackage{graphicx}  
\usepackage{xcolor}

\begin{document}

\begin{frontmatter}



\title{Cybersecurity Practices in Industrial Automation and Control Systems: a Study of Croatian System Integrators}


\author[1]{Filip Katulić} 
\author[1]{Damir Sumina} 
\author[1]{Stjepan Groš} 
\author[2]{Tena Velki} 

\affiliation[1]{organization={University of Zagreb Faculty of Electrical Engineering and Computing},
            addressline={Unska 3}, 
            city={Zagreb},
            postcode={10000}, 
            country={Croatia}}

\affiliation[2]{organization={Josip Juraj Strossmayer University of Osijek Faculty of Education},
            addressline={Cara Hadrijana 10}, 
            city={Osijek},
            postcode={31000}, 
            country={Croatia}}
\begin{abstract}
Given the increasing frequency of industrial cybersecurity incidents, governments, industry, and the research community are all working to find ways to protect industrial automation and control systems (IACSs). A key regulatory effort in the European context is the EU's Network and Information Security Directive 2 (NIS2), which aims to strengthen cybersecurity requirements and improve resilience across critical and important sectors. Nevertheless, translating cybersecurity requirements into practical engineering remains challenging within IACS environments. 
This paper explores how Croatian IACS system integrators perceive cybersecurity requirements, implementation challenges, and current practices in the industrial environments in which they operate. The findings are based on three focus groups with system integrators working with clients in the energy, chemical and pharmaceutical, and food and beverage sectors. Participants reported limited cybersecurity awareness among IACS owners, as well as perceived differences in practices between smaller and larger companies and between domestic and foreign firms. Participants were divided in their expectations regarding regulatory pressure: while some believed it could improve cybersecurity, others expected only minimal compliance focused on cost reduction. The results of this study provide a basis for developing more practical cybersecurity practices and offer a deeper understanding of the key challenges faced by IACS system integrators and owners when dealing with cybersecurity.
\end{abstract}


\end{frontmatter}



\section{Introduction}
\label{Introduction}

Critical infrastructure (CI) encompasses a range of systems and services that are vital for the stability and well-being of modern societies. Industrial automation and control systems (IACSs), including programmable logic controllers (PLCs) and supervisory control and data acquisition (SCADA) systems, are essential for ensuring that CIs operate safely and reliably. However, much of the CI still relies on legacy IACS equipment that was developed without cybersecurity in mind \cite{Knapp2015}. Such systems commonly rely on \textit{security through obscurity}, a strategy that conceals internal system details, enabling advanced persistent threats (APTs) to exploit existing vulnerabilities \cite{Cambadedes2011}. 

The early days of the 2022 Russo-Ukrainian conflict demonstrated that modern conflicts are not limited to conventional warfare. During this period, Russian state-sponsored APTs deployed IACS-specific malware targeting power grid systems \cite{industroyerv2}, \cite{Frostygoop2024}. In 2023, Danish CI was a victim of a coordinated cyberattack in which an APT group gained access to some of the IACSs managing the country's energy sector \cite{SektorCERT2023}. That same year, an Iranian-backed APT targeted PLCs and other IACS equipment commonly used in water supply systems, compromising their normal operation \cite{CISA2023}. This group remains active, reportedly gathering credential information related to CI in the U.S. \cite{CISA2024}. Most recently, in 2024, the Cybersecurity and Infrastructure Security Agency (CISA) reported that APTs sponsored by the People's Republic of China had compromised multiple CI sectors in the U.S., aiming to disrupt their operations in the event of a potential conflict \cite{CISA2025}. 

In response to such threats, the European Union has introduced stronger cybersecurity obligations through the Network and Information Security Directive 2 (NIS2), which expands the scope of regulated entities and introduces stricter expectations for how cybersecurity risks are managed and reported, as well as how responsibility is assigned at the organizational level \cite{EU2022NIS2}. However, how regulatory requirements translate into practical engineering and operational decisions in IACS environments is not yet well understood, and the factors that shape this process in practice remain an open question. Existing literature attributes these challenges to several factors, including legacy equipment \cite{Katulic2023}, limited cybersecurity awareness \cite{Lowawareness}, knowledge gaps \cite{Norway2024}, resource constraints \cite{ASGHAR2019}, and the time required for organizations to adapt to new regulatory expectations.

Most of these assumptions are based on secondary sources of data, such as earlier published papers or anecdotal observations, and have not been systematically validated. One way of gaining a clearer understanding is through primary research using qualitative methods, such as interviews, focus groups, and surveys, which provide firsthand information on cybersecurity practices and challenges in IACSs \cite{Norway2024}.

In this study, we explore IACS cybersecurity implementation from the perspective of Croatian system integrators. System integrators are relevant because they are involved throughout the project lifecycle, from design and configuration to maintenance and communication with asset owners. Their position also allows them to observe how cybersecurity requirements are interpreted and applied in practice. We conducted three focus groups, each consisting of three to four participants, to examine how they deal with cybersecurity requirements, how they respond to incidents, which measures they implement or encounter, and how they view regulatory pressure. The following main research questions are discussed in the paper:

\begin{itemize}
\item How do IACS system integrators perceive cybersecurity requirements received from asset owners?
\item How do system integrators describe current practices for monitoring, responding to, and assigning responsibility for cybersecurity incidents?
\item Which technical cybersecurity measures do system integrators report implementing or encountering in IACS environments?
\item How do system integrators expect regulatory requirements, particularly NIS2, to influence cybersecurity implementation?
\end{itemize}

The analysis identifies several recurring themes in participants' responses. Participants described cybersecurity awareness among asset owners as uneven and often limited, while suggesting that cybersecurity maturity may vary depending on company size, sector, ownership structure, and exposure to foreign clients or suppliers. They also reported unclear responsibility boundaries during cybersecurity incidents, particularly where maintenance contracts do not explicitly define monitoring and incident response obligations. Incident handling was commonly described as reactive, with priority given to restoring system availability rather than conducting detailed root-cause analysis. In relation to technical measures, participants reported that general IT-oriented controls, such as firewalls and VPNs, were more commonly encountered than IACS-specific cybersecurity solutions. Finally, participants expressed mixed expectations regarding regulation: some expected NIS2 to increase cybersecurity investment, while others anticipated formal or minimal compliance with limited practical change.

The novelty of this study lies in examining IACS cybersecurity implementation from the perspective of system integrators. Their position between asset owners, vendors, and operational environments allows them to observe the gap between cybersecurity requirements and actual engineering practice. This perspective helps explain why implementation difficulties often arise from organizational and contractual conditions, rather than from technical limitations alone.

The remainder of this paper is structured as follows. In Section \ref{sec:background}, the related work is presented. Section \ref{sec:Method} describes the focus group methodology used in the study. Section \ref{sec:sec4} presents the results, including each of the identified themes. Finally, Section \ref{sec:sec5} provides a discussion, while the conclusion and future work are given in Section \ref{sec:sec6}.  

\section{Background and Related Work} \label{sec:background}

Over the past decade, industrial cybersecurity has received significant attention from the research community \cite{BHAMARE2020101677}. Various studies have analyzed cybersecurity vulnerabilities, challenges, and best practices in IACSs, relying primarily on secondary data sources such as existing research papers and standards \cite{Lowawareness}, \cite{ASGHAR2019}, \cite{BHAMARE2020101677}, \cite{IEC62443-1-1_2007}. This literature is useful because it summarizes known weaknesses and available protections, but it says less about how cybersecurity is actually managed during projects and maintenance. This is important because a gap often exists between formal cybersecurity requirements and practical implementation, especially when procedures are difficult to apply in real work settings \cite{Security_culture}.

Primary research offers a different perspective. Surveys, focus groups and interviews can reveal how practitioners understand risks, what they do in practice, and what prevents stronger cybersecurity from being implemented. Qualitative methods are particularly useful when the focus is on human judgment, incident response, daily work routines, or communication between different professional groups \cite{HumanfactorsInCyber}, \cite{QualitativeMethodsInCyber}. Before continuing, it should be noted that in the literature, the terms OT, ICS, and IACS are often used interchangeably. To remain consistent with the ISA/IEC 62443 terminology, this paper uses the term IACS \cite{IEC62443-1-1_2007}.

Several empirical studies have examined cybersecurity challenges in CI and IACSs. In \cite{dnv2025cyber}, the authors conducted a survey of 375 employees in energy CI about challenges in achieving cyber resilience. Their results show that lack of skills, supply chain complexity, and general unpreparedness pose the most significant threats to organizational cybersecurity. The same authors also conducted interviews with industrial cybersecurity experts, confirming the survey results. Furthermore, the authors in \cite{SANSICS}, also conducted a survey across multiple CI sectors, investigating the use of SANS Five Cybersecurity Critical Controls, a set of measures to lower risks and ensure CI cybersecurity. Their results show that low cybersecurity awareness, along with a tendency for company budgets to prioritize technology over personnel, represents the greatest risk to cybersecurity effectiveness. These studies point to a simple issue: cybersecurity is not weakened only by missing technology, but also by limited capacity to implement and maintain it.

A more detailed view of current IACS cybersecurity challenges appears in \cite{Norway2024}, where interviews were conducted with IACS owners from ten CI-related companies. The study examined supply chain issues, vulnerability management, and the use of testbeds and digital twins. It reported limited visibility into supply chain risks, weak cybersecurity attention in procurement, difficulties with asset inventories, and resource constraints. One point from that study is especially relevant: it explicitly notes that interviews with system integrators and equipment manufacturers could provide another perspective, since asset-owner interviews capture only part of the implementation process \cite{Norway2024}.

Supply chain issues also appear in \cite{supply_interview}, where interviews with smaller energy distribution companies were used to identify security gaps and propose requirements for improving supply chain security. The paper shows that cybersecurity in industrial environments often depends on relationships between organizations, not only on what the asset owner does internally. This is relevant for system integrators, who may be involved in project delivery and later maintenance, while having limited influence over budgets, internal policies, or final operational decisions.

The role of system integrators is also reflected in standards. The IEC 62443 series separates the asset owner role from the role of service providers during integration and maintenance. IEC 62443-2-4 describes security program requirements for IACS service providers and connects them directly to integration and maintenance activities \cite{IEC62443-2-4_2023}. This is important because several problems discussed in empirical studies occur in this area. If monitoring, incident response, or maintenance obligations are not clearly defined, responsibility can become unclear after handover. In such cases, weak implementation is not only a technical issue, but also a question of how work is assigned and contracted.

Other empirical work has examined organizational and cultural barriers in greater detail. In \cite{Security_culture}, interviews with professionals from IACS sectors revealed that cybersecurity tasks are often added to existing roles without adequate training. The same study reported governance issues, communication difficulties between functions, and a lack of IACS cybersecurity expertise. It also discussed the influence of external consultants and security vendors on security culture. This is relevant to the present work because system integrators are also external actors, and their impact on implementation may be greater than it initially appears. Recent studies further suggest that some implementation problems are related to usability. In \cite{Nunes2024UserCentredICS}, interviews with experienced IACS security experts indicated that many cybersecurity solutions are not deployed because they are based on assumptions that do not align with actual end-user needs in IACS environments. In \cite{Li2023UsableSecurityICS}, a participatory study showed that security mechanisms such as authentication and access control can still be misconfigured when legacy systems offer limited support or feedback to users.

These findings are relevant because they suggest that low adoption of IACS-specific security measures may result from practical challenges in deployment and operation, not just a lack of awareness or budget. A similar issue arises in work on security-by-design. In \cite{Elmarkez2025SecurityByDesignICS}, a systematic mapping study found that many approaches focus on design and early security analysis, while the integration of mitigation measures into real-world deployment is less developed. The same study also noted that conflicts between security requirements and other system needs are not always well managed. This fits well with IACS environments, where safety, availability, and maintenance windows often restrict how quickly changes can be implemented. Broader research on the IACS security landscape has also emphasized that industrial constraints and cyber-physical coupling make it difficult to apply enterprise IT approaches directly \cite{McLaughlin2016ICSLandscape}, \cite{NIST80082r3}. Overall, existing empirical work provides a comprehensive view of many IACS cybersecurity challenges, but mainly from the perspectives of asset owners, cybersecurity professionals, or broader CI populations. Less is known about system integrators, even though they operate close to the point where requirements are translated into configured systems and subsequent maintenance practices.

A related methodological issue is the limited use of focus groups in IACS cybersecurity research. Interviews remain the most common qualitative method in cybersecurity research, as noted in \cite{QualitativeMethodsInCyber}. Focus groups are less frequently used, but the discussion in \cite{LessonsLearned} demonstrates that they can be useful for comparing professional experiences and observing how participants respond to each other's views. This is relevant for the present study because many implementation problems are not only individual experiences but also shared patterns that emerge across projects and clients. At the same time, focus groups have limitations that are particularly significant in cybersecurity research. The study in \cite{LessonsLearned} notes that participants may be reluctant to share sensitive information about vulnerabilities, incidents, or internal procedures. Similar issues are reported in \cite{BadaAfrica}, where differences in expertise and hierarchy affected openness and comparability across focus groups. In \cite{API}, workplace hierarchy also made some participants less willing to express critical opinions.

Nonetheless, focus groups have already been used in several cybersecurity-related contexts. In \cite{cybersecurity_awareness_focus_group}, they were used to examine cybersecurity awareness across different target groups, including employees working in CI-related companies. The study identified practical limitations related to funding, coordination, and the preparation of awareness materials. In \cite{CSIRT}, focus groups with Computer Security Incident Response Team (CSIRT) experts showed that participants were sometimes reluctant to share operational details, which affected the quality and comparability of the collected data.

Based on the reviewed work, two gaps are relevant for this paper. First, empirical IACS cybersecurity research has focused more on asset owners and cybersecurity practitioners than on system integrators, even though standards explicitly recognize the service provider role during integration and maintenance \cite{IEC62443-2-4_2023}. Second, focus groups remain relatively uncommon in IACS cybersecurity research, even though they can reveal how practitioners compare shared problems and differences in experience \cite{LessonsLearned}, \cite{QualitativeMethodsInCyber}. This study addresses these gaps through focus groups with Croatian IACS system integrators. It examines how they describe cybersecurity requirements, incident handling, implemented measures, and expectations related to NIS2-driven regulatory pressure. Recent European guidance shows that NIS2 implementation requires translating legal obligations into concrete measures and evidence, making the gap between regulation and practice even more relevant \cite{ENISA_NIS2_Guidance}.
\section{Research Method}
\label{sec:Method}
This study employs qualitative methods, specifically focus groups, to examine cybersecurity practices and challenges among IACS system integrators. Qualitative methods are appropriate when the goal is to understand human judgment, lived experience, and practice rather than to measure variables or test hypotheses \cite{HumanfactorsInCyber}, \cite{QualitativeMethodsInCyber}. Focus groups are particularly suited to this study because many of the challenges under investigation are not purely individual experiences but shared patterns that emerge across projects and clients, making group interaction a productive way to surface them \cite{LessonsLearned}, \cite{Morgan1997-ot}.

In general, focus groups are a qualitative research method that involves gathering a group of participants for a moderated discussion based on their perceptions and experiences of a researcher-defined topic \cite{LessonsLearned}. Focus group sizes typically range from four to a maximum of twelve participants, depending on the topic and the target group \cite{Carlsen2011_wc}. However, an interesting finding is presented in \cite{4_vs_8_focus}, where the authors concluded that conducting two focus groups of four participants yielded more information than a single group of eight participants. This suggests that increasing the number of focus groups while reducing the number of participants per group is a reasonable approach when the overall sample size is small. Regarding the required number of focus groups, most recommendations on sample size refer to the point of  \textit{data saturation}, where researchers continue conducting focus groups until no new or relevant data emerge within a given question category \cite{Carlsen2011_wc}. Focus groups allow researchers to gain a range of perspectives, as the interaction between participants often brings up topics that would be difficult to uncover when conducting individual interviews \cite{Morgan1997-ot}. Finally, focus groups are also recommended to complement and interpret data collected through other qualitative research methods \cite{Lambert2008-dl}, e.g., questionnaires \cite{focus_quest}.
\subsection{Focus Group Design}
\label{subsec1}
To address the gaps identified in previous research and gain a broader perspective on the cybersecurity practices of Croatian system integrators in IACSs, we conducted three focus groups with three to four participants each from various CI-related companies. When designing the focus group questions, we followed the principles outlined in \cite{Krueger2002DesigningAC}, which suggest the use of open-ended questions for facilitating the discussion. The questions were developed through a brainstorming process among the authors of this paper, aiming to understand cybersecurity practices of system integrators in IACSs. Specifically, the idea was to gain insights into their perception of cybersecurity maturity in their respective sectors, and to gather information on the cybersecurity measures they implemented or encountered in daily work. Additionally, we wanted to compare the cybersecurity practices of system integrators with those of IACS asset owners, which were analyzed by the authors of \cite{Norway2024}. Following this phase, a structured focus group script was created, which is provided later in the section. 

The focus group sessions were structured into three phases. The introductory phase involved outlining the goals of the discussion, obtaining written informed consent from participants, and explaining the measures taken to protect participant confidentiality, as any potential disclosures could have implications for their companies. During the discussion phase, the main research questions were discussed, which are provided later in the section. The third phase involved a wrap-up, during which we addressed any remaining questions regarding the focus group research.

\subsection{Participants}
\label{subsec2}
We invited several established Croatian companies from different industry sectors with experience in both domestic and foreign IACS projects, requesting that they send representatives from both the management and engineering roles. This approach was intended to capture a broad range of perspectives, from strategic decision-making to technical implementation. According to the size categorization given in \cite{EU2003}, all but one of the companies included in the focus group study are classified as small and medium-sized enterprises (SMEs), while one company falls into the large enterprise category. Given the size of the Croatian market, the number of companies operating as IACS system integrators, the market share of the companies interviewed, and the fact that these companies also operate abroad, the sample size can be considered appropriate for an exploratory qualitative study. All of the participating companies operate not only within Croatia but also deliver IACS integration projects across the EU and other international markets, which further supports the relevance of their perspectives. More on this issue is given in the discussion section. The study was conducted in accordance with institutional ethical requirements. Participants provided written informed consent, and all company and participant identifiers were anonymized.

The first group involved four employees from a system integrator company in the energy sector: a company director, an IACS cybersecurity project manager, and two IACS engineers. The second focus group included three employees from two companies working in the chemical and energy sectors. The first company was represented by a director and an IACS engineer, while the other company was represented by a director. The third focus group involved four employees from a company working with the food and beverage industry, consisting of the company director, an IT cybersecurity specialist, and two IACS engineers. Further details of the participants are provided in Table \ref{tab1}. 

Each focus group session lasted between 90 and 120 minutes, during which all the research questions were discussed. The first and third sessions were conducted at the respective companies, while the second session took place at our institution. Participants from both managerial and engineering roles had over five years of professional experience in IACS design, integration, or management.

\begin{table} [!b] 
\caption{\textbf{Participants grouped by focus group, company/sector, and role}}
\centering 
\setlength{\tabcolsep}{3pt}
\resizebox{0.9\columnwidth}{!}{
\begin{tabular}{l l l}
\hline
\textbf{Group No.} & \textbf{Company / Sector} & \textbf{Role} \\ \hline
1. & A / Energy & Director \\
  &            & IACS cybersecurity project manager \\
  &            & IACS engineer \\
  &            & IACS engineer \\ \hline
2. & B / Energy & Director \\ \cline{2-3}
  & C / Chemical and & Director \\
  &    pharmaceutical        & IACS engineer \\ \hline
3. & D / Food and beverage & Director \\
  &            & IT cybersecurity specialist \\
  &            & IACS engineer \\
  &            & IACS engineer \\ \hline
\end{tabular}
}
\label{tab1}
\end{table}

\subsection{Instruments}
\label{subsec3}

The following main research questions were discussed: 

\textit{Category A – Cybersecurity awareness}:
What cybersecurity requirements do you receive from IACS owners?
How do you communicate with owners who do not prioritize cybersecurity?

\textit{Category B – Response to threats}:
How do you currently monitor and respond to cybersecurity threats in IACSs?
Are there any responsibility issues when it comes to cybersecurity incidents?

\textit{Category C – Technical measures}:
Which technical cybersecurity measures do you implement or find in IACSs?
Do you implement IACS-specific cybersecurity measures?

\textit{Category D - Legislation}:
Do you believe that regulation will compel the IACS owners to improve their cybersecurity posture?

\subsection{Data analysis}

The data analysis began with formatting and annotating the collected transcripts, as well as reviewing manually collected transcripts and comments. All focus group discussions were transcribed during the sessions using an offline version of  \textit{Live Transcribe} tool \cite{LiveTranscribe}, supplemented with manual transcriptions and comments from the interviewers. The formatted transcripts were then imported into \textit{Taguette}, an open-source software tool designed for qualitative data analysis \cite{Taguette}. To protect participant privacy, we used a self-hosted version of the software. Our approach to data analysis closely followed the method described in \cite{Braun01012006}, where the authors defined thematic analysis as a process of identifying repeated patterns of meaning within the data. This method was chosen due to its widespread use and flexibility, allowing researchers with limited experience to conduct qualitative research \cite{Braun01012006}. 

The method consists of six steps, in which the first step involves familiarization with the collected data by reviewing the collected transcripts and identifying recurring themes across focus groups. The second step focuses on generating an initial set of codes, which are descriptive labels used to organize data into meaningful groups \cite{Tuckett01082005}. For example, \textit{virtual private network} (VPN) was used as a code, as multiple participants mentioned encountering VPNs in IACSs. In the third step, initial codes were assigned to preliminary created themes, depending on the context. For instance, the previously mentioned \textit{VPN} was assigned under the theme \textit{Technical measures – IT measures}, as VPNs were one of the IT cybersecurity measures found within IACSs. Other similar measures were also grouped under the same theme, enabling a comparative analysis of the occurrence of specific technologies and practices within IACSs. 

The fourth and fifth step involved reviewing and refining the themes by discarding those with insufficient supporting data and breaking down larger themes into smaller, more specific ones. Because the study used thematic analysis in an interpretive manner, a formal inter-coder reliability coefficient was not treated as a necessary quality criterion. Inter-coder reliability can be useful in codebook-driven analyses, but it is debated in interpretive qualitative research, where coding decisions form part of the analytical process rather than independent measurements of fixed categories \cite{BraunClarke2021Quality},\cite{OConnorJoffe2020}. To strengthen rigor, the transcripts were coded independently by the researchers, coding disagreements were discussed, and the final themes were agreed through consensus. It should also be noted that the focus groups were conducted in Croatian. Quotations used in the manuscript were translated into English by the authors and checked for consistency with the original meaning.

\begin{figure} [t!]
\centering 
\includegraphics[width=0.7\columnwidth,trim=0 0 0 0, clip]{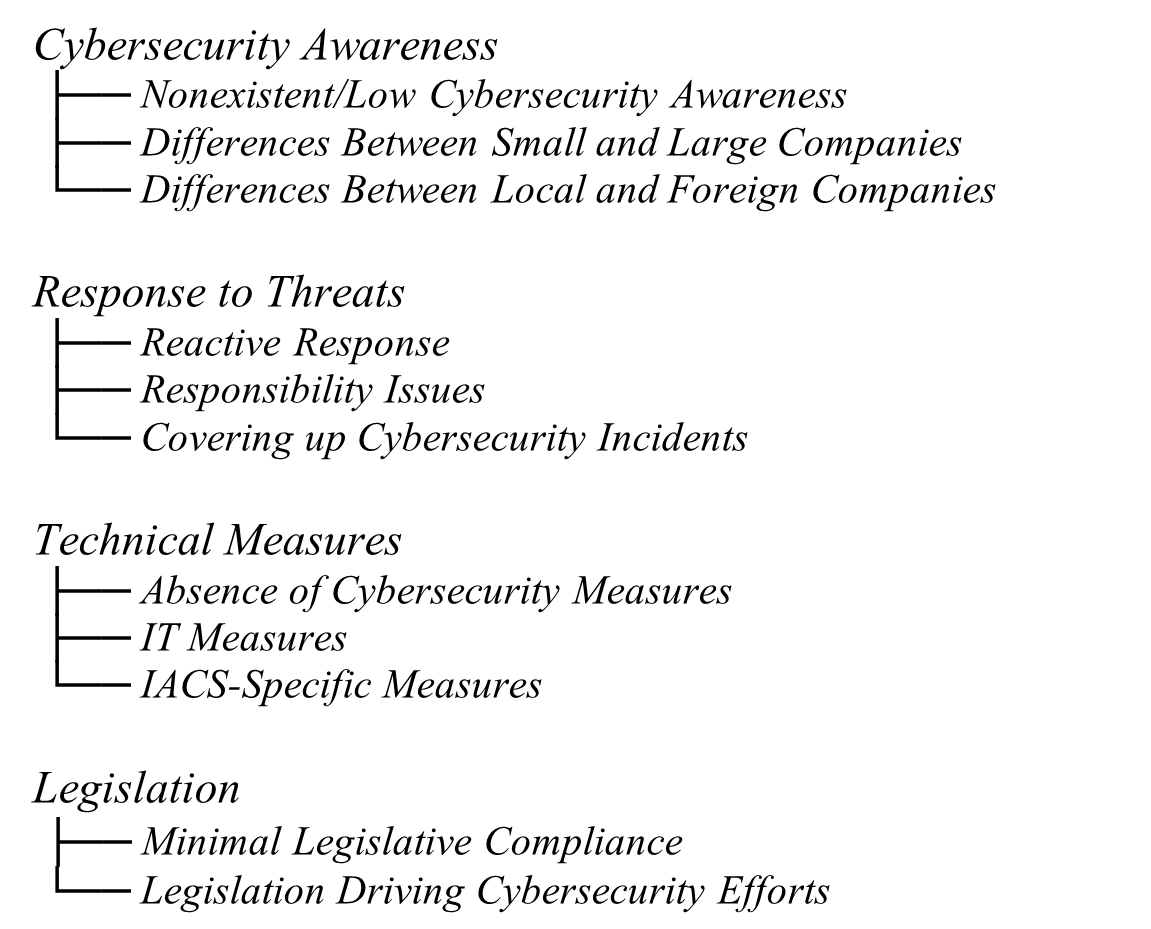}
\caption{Identified themes and subthemes of responses}
\label{fig1}
\end{figure}

Our study identified four main themes and their respective categories of responses, as presented in Fig \ref{fig1}. A detailed analysis of each identified theme is provided in the following section. Finally, our research ended in a report, in which we identified the main findings of the study. 

\section{Results}
\label{sec:sec4}
The analysis resulted in four main themes: cybersecurity awareness, response to threats, technical measures, and regulatory pressure. Each theme contained several subthemes that describe how participants experience cybersecurity implementation in the client environments in which they work. For clarity, quotations are referenced in the text as (\textit{Q1}), (\textit{Q2}), etc., while the full illustrative quotations are provided in the Appendix. The results should be read as qualitative findings based on participants' accounts, rather than as measurements of prevalence across the wider IACS sector.

\subsection{Cybersecurity Awareness}
Three subthemes emerged within the cybersecurity awareness theme: nonexistent or low awareness, differences between small and large companies, and differences between domestic and foreign companies. The first subtheme was the most frequently mentioned, recurring across all three focus groups and across participants from both managerial and engineering roles.
\subsubsection{Nonexistent/Low Cybersecurity Awareness}
Across all three focus groups, participants consistently described cybersecurity awareness among IACS owners as low or effectively absent. As key issues, the participants reported a general lack of cybersecurity awareness among IACS owners (Q1), management (Q2), and employees working in IACSs (Q3). Participants noted that IACS owners tend to prioritize functional safety and regulatory compliance, such as hazard and operability studies (HAZOP), while treating cybersecurity as a secondary concern (Q4). A recurring explanation was that many owners do not consider themselves likely targets, thinking that larger or more prominent organizations present more attractive opportunities for attackers (Q5). Beyond this, participants described a broadly passive attitude, with many companies waiting to observe what others do before taking any action themselves (Q6). What is also troubling is that cybersecurity considerations are often excluded from financial plans, even when developing greenfield IACSs (Q7).

\subsubsection{Differences Between Small and Large Companies}
Within the same topic, a discussion about differences between small and large companies emerged. It should be noted that we did not apply strict criteria regarding the number of employees, such as those defined in \cite{EU2003} for categorizing SMEs and large enterprises. Instead, we followed the participants’ own definitions, where small companies were typically seen as having a smaller number of staff, flatter organizational structures, and fewer internal resources, while large companies were described as having more formalized processes, larger teams, and greater access to funding and infrastructure. Participants identified sectors such as the pharmaceutical industry and parts of the energy sector as showing relatively higher cybersecurity awareness, while water and wastewater systems were described as an example of the opposite extreme (Q8). Such conclusions are based on the fact that large companies usually have an educated staff, such as IACS engineers and specialists, while smaller companies cannot afford to invest in or hire such expertise (Q9). It is generally simpler to work with large clients that have specialized cybersecurity teams, as IACS system integrators can openly discuss which cybersecurity solutions could be better fit for their IACS (Q10). Although some larger companies have had IT cybersecurity programs for years, it was not until recently that they started introducing cybersecurity requirements when working with IACS system integrators and their suppliers (Q11). However, despite some companies having invested significant financial, time, and organizational resources, the level of implementation of cybersecurity solutions remains low (Q12).

\subsubsection{Differences Between Domestic and Foreign Companies}
 A third subtheme concerned differences in cybersecurity practices between domestic and foreign companies, which emerged in two of the three focus groups. Participants noted that foreign companies, particularly those from regions with higher cybersecurity maturity, integrated cybersecurity solutions in IACSs they later maintained (Q13). Furthermore, when maintaining IACSs in those countries, strict cybersecurity measures were enforced, such as screen recording and time-limited access, which could potentially increase response time in urgent situations (Q14). In contrast, the situation in developing countries appears significantly different, with many IACS employees lacking even a basic understanding of cybersecurity principles (Q15).

\subsection{Response to Threats}
\label{subsec5}
Three subthemes emerged within the response to threats theme: reactive incident response, responsibility issues, and covering up cybersecurity incidents. Discussion of this theme was concentrated among participants in engineering roles, who drew on direct experience of handling incidents in client environments. Notably, participants across all three groups described limited experience with systematic threat response, reflecting the low cybersecurity maturity of the environments in which they work.

\subsubsection{Reactive Response}
Participants across all three focus groups described incident response as predominantly reactive. When a cybersecurity incident is identified, IACS owners typically contact the system integrator responsible for maintenance, with the immediate priority being restoration of system availability rather than investigation (Q16). Participants noted that, as a result, they usually do not conduct in-depth investigations into the root cause of the incident and only escalate the issue to external services or companies if they are unable to resolve it themselves (Q17). In the event of an incident, they typically restore from a system backup or a system image, which usually resolves the issue (Q18).

\subsubsection{Responsibility Issues}
When it comes to responsibility for cybersecurity, most participants viewed IACS owners as primarily responsible for cybersecurity after system handover, as contracts typically do not define cybersecurity obligations (Q19). While some companies provide cybersecurity monitoring, these services are primarily advisory (Q20). Similarly, maintenance contracts place full responsibility on the IACS owners (Q21). One participant reported handling multiple cybersecurity incidents caused by IACS owners introducing infected equipment into their systems (Q22). Additionally, some IACSs rely on external cybersecurity providers for Security Operations Center (SOC) services, however, the high number of false-positive and false-negative alarms makes these services challenging to work with (Q23). 

\subsubsection{Covering up Cybersecurity Incidents}
A third subtheme, which emerged in two of the three focus groups, concerned the deliberate concealment of cybersecurity incidents by IACS owners. Participants suggested that many companies do not disclose cybersecurity incidents, which could be a major contributing factor to recurring security breaches (Q24). A concrete example was described in which one IACS owner insisted on justifying the costs of replacing IACS equipment infected with ransomware by attributing them to the installation of 100 meters of cable, attempting to cover up the incident (Q25).
\subsection{Technical Measures}
\label{subsec6}
Three subthemes emerged within the technical measures theme: absence of cybersecurity measures, general IT measures, and IACS-specific measures. This theme generated the most discussion across all three focus groups, with participants drawing on direct experience of both deploying and encountering measures in client environments. The overall picture that emerged was one of limited technical protection, with IT-oriented controls representing the most commonly encountered measures and IACS-specific solutions remaining rare.

\subsubsection{Absence of Cybersecurity Measures}
The most consistently recurring observation across all three groups was that the majority of IACSs participants work with have minimal or no technical cybersecurity measures in place (Q26). Some of these systems even have flat network architectures, despite being physically distributed, meaning a threat actor could potentially access the entire IACS through a single access point. Furthermore, cybersecurity and availability requirements are almost never balanced, as cybersecurity requirements are often not considered at all (Q27). One participant highlighted the difference between large and small companies, noting that while larger companies may attempt to implement cybersecurity measures, smaller companies almost never do (Q28).

\subsubsection{IT Measures}
Where technical measures were present, participants described them as predominantly general IT security controls rather than IACS-tailored solutions. Table \ref{tab2}. presents the occurrence of various IT measures in the collected responses. As can be seen, firewalls are the most reported IT measure, followed by VPNs and intrusion detection systems (IDSs). However, several participants stressed that the effectiveness of implemented measures depends on proper configuration, which is often not the case (Q29). Furthermore, IDSs are commonly present at the IT level, rather than in IACS networks (Q30). In general, companies usually start with basic cybersecurity measures, such as passwords, network logging, and general system hardening (Q31). 

\begin{table} [!b]
\centering 
\small
\caption{\textbf{IT cybersecurity measures occurrence in the participants’ answers}}
\resizebox{0.8\columnwidth}{!}{
\setlength{\tabcolsep}{3pt}
\begin{tabular}{l l}
\hline
\textbf{No. of mentions} & \textbf{Security measure} \\ \hline
5 & Firewall \\ \hline
4 & VPN \\ \hline
3  & IDS \\ \hline
1  &    VLAN (Virtual LAN) \\ \hline
1 & PAM (Privileged Access Management)\\  \hline
1  &  SOC (Security operations center) \\  \hline
\end{tabular}
\label{tab2}
}
\end{table}
\subsubsection{IACS-Specific Measures}
IACS-specific cybersecurity measures were the least frequently encountered subtheme across all three groups. Before asking this question, we explained to participants that IACS-specific measures refer to cybersecurity measures designed specifically for IACSs, rather than general IT security controls. The question was intentionally phrased broadly to encourage a larger number of responses, but we received relatively few answers, likely reflecting the low cybersecurity maturity of the clients for which the system integrators provide services. Several participants mentioned firewalls for IACS communication protocols as a technology they had implemented in multiple IACSs (Q32). Additionally, some participants noted the use of industrial remote access solutions, encompassing both hardware and software components (Q33). One participant indicated that user management through Active Directory (AD) is sometimes implemented separately in IACS and IT networks (Q34).

\subsection{Legislation}
\label{subsec7}
Finally, two main views emerged regarding how legislation impacts cybersecurity in IACSs: one suggesting that minimal effort will be made to ensure compliance, and the other indicating that regulatory requirements could improve the overall cybersecurity posture within IACSs. Both subthemes appeared across all three focus groups, and in several cases individual participants expressed elements of both views, acknowledging that legislation might force action while doubting the quality of that action.

\subsubsection{Minimal Legislative Compliance}
Participants who expressed skepticism described a pattern they expected to see repeated with NIS2: companies becoming aware of a regulatory obligation and responding by seeking the most efficient path to formal compliance rather than substantive cybersecurity improvement (Q35). Some participants noted that while companies may be aware of regulations, they rarely integrate cybersecurity into procurement (Q36). One participant expressed concern that cybersecurity will be managed by lawyers rather than cybersecurity experts (Q37). Another compared the situation to safety HAZOP analyses, where an IACS owner kept searching for a company that would declare them compliant, despite previous assessments identifying errors and noncompliance (Q38).
\subsubsection{Legislation Driving Cybersecurity Efforts}
Participants who expressed a more optimistic view did not dispute that initial responses would be compliance-driven, but argued that regulatory pressure would nonetheless produce real change over time (Q39). Most of them believe that only the risk of financial penalties will force them to implement the necessary measures (Q40), as cybersecurity awareness is generally low. Furthermore, they identified NIS2 Directive as key regulation in driving the change (Q41), as it introduces individual responsibility in case of an incident. However, some participants noted that due to the short timeline for implementing the NIS2 Directive into national laws across European countries, some companies may take advantage of the market chaos, falsely advertising programs for compliance (Q42). 
\section{Discussion}
\label{sec:sec5}
The low cybersecurity awareness reported by participants across all three groups is consistent with findings from asset-owner focused studies \cite{Norway2024, dnv2025cyber, SANSICS} and with research on organizational barriers in OT environments more broadly \cite{Security_culture}. However, the system integrator perspective adds a dimension that those studies do not capture: awareness gaps among owners have direct consequences for how integrators can operate. When owners do not understand what cybersecurity requires, they cannot specify it in procurement, cannot allocate budget for it, and cannot engage meaningfully with integrators about appropriate solutions. The result, as participants described, is that cybersecurity decisions are effectively delegated to the integrator by default, not through explicit agreement but through the absence of any owner-side direction. Participants also noted that cybersecurity is frequently excluded from financial planning even in greenfield projects, where security measures could most cost-effectively be integrated from the outset \cite{doe2022cie}. Without a contractual requirement, integrators have neither the mandate nor the financial basis to include security measures in their scope. This points to a procurement-level failure that regulation alone may not resolve unless it reaches into how contracts between owners and integrators are structured. Regulatory implementation of NIS2 should therefore explicitly require that cybersecurity specifications appear in tenders and contracts for IACS projects, rather than leaving this to the discretion of individual owners \cite{ENISA_NIS2_Guidance}.
	
Next, a difference in cybersecurity in domestic and foreign companies was identified, implying different levels of cybersecurity maturity. The results are similar to the ones presented in \cite{Creese2021-je}, where the authors quantitatively analyzed the differences in general IT cybersecurity practices across the population of 78 nations, showing that developed countries have a significantly larger cybersecurity maturity stage compared to other countries. The given result could indicate a certain level of local immaturity in cybersecurity practices within the Croatian industrial context and, as such, should be interpreted with caution and not generalized to other European countries.
	
 One of the key topics identified was the difference between small and large companies managing IACSs, with larger companies demonstrating higher cybersecurity awareness and maturity. A similar trend is observed in the IT sector, where SMEs also display lower cybersecurity awareness and maturity \cite{Heidt2019-uw}. With smaller or less mature clients, integrators described adapting to whatever the client would accept, often implementing less than they considered adequate. This suggests that the cybersecurity outcome of a given IACS project depends substantially on the client's organizational maturity, not only on the integrator's technical capability. To address this challenge, some researchers propose lightweight cybersecurity frameworks that help owners to establish a fundamental cybersecurity posture without imposing excessive costs and complexity \cite{lightweight}. It would be beneficial for European countries to consider this difference when transposing the NIS2 directive to their national laws, ensuring that the specific challenges faced by SMEs in securing IACSs are addressed. Additionally, these companies will also require both educational and financial support, as full compliance with regulatory requirements may prove challenging.
	
Furthermore, challenges related to responsibility in the event of cybersecurity incidents were highlighted, as such scenarios are not currently defined in maintenance contracts between IACS owners and system integrators. This issue arises from the previously identified lack of cybersecurity maturity. For example, the authors of \cite{Katulic2023} suggest that IDSs and similar devices that detect cybersecurity incidents in IACSs should be capable of notifying a designated responsible person when malicious behavior is detected. However, our results indicate that neither IACS system integrators nor owners clearly know who the person responsible is. Similar findings were reported in \cite{IDS_Challenges}, where the authors conducted nine interviews with IT cybersecurity practitioners discussing challenges when implementing IDSs in IT systems. Addressing this lack of clarity requires contractual measures, with contracts explicitly defining roles and responsibilities for monitoring, detection, reporting, and mitigation of cybersecurity incidents. IEC 62443-2-4 already provides a framework for security program requirements for IACS service providers \cite{IEC62443-2-4_2023}, but participants' responses suggest its obligations are rarely reflected in actual maintenance contracts. Industry associations and regulators should consider developing sector-specific contract templates that translate these obligations into language owners and integrators can apply directly.
	
The reactive and non-systematic incident response practices described by participants suggest limited organizational experience with structured cybersecurity incident handling in IACS environments. This issue could potentially be mitigated with investments in personnel training. However, such investments could pose challenges for system integrators working with multiple equipment manufacturers, as they would need to allocate significant resources to cover a variety of platforms. Consequently, this could drive IACS system integrators to work with a smaller number of equipment manufacturers, potentially reducing market diversity and competition. A more practical approach would be to define vendor-neutral incident response procedures and escalation paths, while relying on vendor-specific expertise only when deeper technical recovery is required. Such procedures should clarify how incidents are identified, who is contacted, how evidence is preserved, and how restoration is coordinated between the asset owner, system integrator, external SOC provider, and equipment vendor.
	
 When it comes to cybersecurity equipment implemented in IACSs, participants reported that the majority of systems they work with have minimal or no technical cybersecurity measures in place. According to the ISA/IEC 62443 industrial cybersecurity standard, the implementation of technical measures, alongside non-technical measures such as policies and procedures, is a key strategy for mitigating identified cybersecurity risks within IACSs \cite{IEC62443-1-1_2007}. However, the use of technical cybersecurity measures that require a high level of knowledge and disrupt standard workflows of IACS employees poses a significant challenge, as noted in \cite{Katulic_Devising}. The solutions must be usable, meaning that the burden on users when deploying, maintaining, and operating such equipment should be reasonably low. While it is expected that introducing cybersecurity into previously unsecured IACSs will cause some disruption and require adjustments to existing workflows, the issue of usability is well documented in IT systems and is likely to be even more significant in IACS environments \cite{Katulic_Devising}. The results suggest that cybersecurity solutions must be specifically designed for IACSs, with IACS equipment manufacturers prioritizing usability and backward compatibility with existing infrastructure. As reported by focus group participants, the adoption of IACS-specific cybersecurity equipment is very limited, and in many cases, virtually nonexistent. Such findings may indicate that users' experiences with current technologies have been unsatisfactory or impractical. Future work will focus on examining this issue in greater detail.

 Finally, the difference in views regarding the impact of legislative regulations on IACS cybersecurity posture is also indicative. Around half of participants' answers reported that regulation will drive IACS owners to invest in cybersecurity. Most of the participants identified the risk of financial penalty as the initial drive for investments. Their perspective is in line with the work presented in \cite{Financial_penalties}, where the author identified legal sanctions as one of the motivators for initiating cybersecurity improvements. 
 
 Approximately half of the responses indicated that only minimal compliance with the legislation would be achieved. This suggests that while the compliance level must remain achievable, particularly for SMEs, it must also provide sufficient protection against identified threats. In an unpublished survey reported by the authors of \cite{KNOWLES201552}, which focused on the North American Electric Reliability Corporation Critical Infrastructure Protection (NERC CIP) standards, 77\% of respondents agreed that the standard is necessary for ensuring cybersecurity. However, 70\% believed that the standard alone is insufficient. This actually highlights a broader issue: meeting basic compliance does not guarantee sufficient cybersecurity level. It is reasonable to expect such an issue with relatively new NIS2 targeting a large set of industries and companies, where setting a minimum compliance level that is both effective and achievable could be a major challenge.

 \subsection{Limitations}
Regarding the limitations of this study, we followed the guidelines provided in \cite{limit_focus} when determining the number of groups to conduct. The authors concluded that approximately 80\% of data saturation is typically reached after two or three focus groups, and around 90\% after three to six focus groups. As earlier stressed, data saturation refers to the point at which no new information emerges from additional data collection \cite{Carlsen2011_wc}. The recurrence of the same themes across focus groups suggests that a substantial degree of thematic saturation may have been reached, although additional focus groups could reveal further sector-specific variation.

Nevertheless, one clear limitation of this study is that we conducted only three focus groups, which is a small number and may limit the range of perspectives captured. However, the participants in these groups had very specific knowledge relevant to the topic, and the thematic focus of the study was narrow, which makes the findings still relevant \cite{small_focus_group}. Recruiting additional participants was challenging due to the small size of the available pool of experts. In the future, we plan to expand the participant base, including international experts, potentially by using \textit{virtual} focus groups. Although virtual focus groups offer many benefits, particularly in recruiting participants, they also pose challenges, notably the increased risk of disclosing sensitive information, which may make participants more reluctant to share confidential details \cite{LessonsLearned}. Conducting additional focus groups, particularly within the same sectors, could enable more robust comparisons of practices across companies. Similarly to the authors of \cite{API}, our focus groups consisted exclusively of male participants. This is largely due to the difficulty of recruiting female engineers with relevant experience, a challenge that was even greater in the context of Croatian IACS system integrators. While three of the companies that participated in the focus group discussions do employ female engineers, these individuals either do not work directly with IACS equipment or lacked the experience level met by the other participants, and were therefore not included in the discussion.

Furthermore, one of our conducted focus groups only had three participants, which is commonly known as a triad focus group. Some researchers argue that focus groups should have at least four participants to ensure focus group diversity, while others stress that focus group should be conducted even if a lower number of participants show up to the focus group, as long as their expertise in the topic is significant \cite{small_focus_group}. Based on that, we proceeded with the group and included the results in the research, as their responses were relevant to the research objectives defined. Moreover, all the companies represented in the focus groups work with multiple different IACS owners, allowing us to capture a broad perspective across different sectors.

There are several potential limitations in conducting qualitative research, including issues related to power dynamics between participants. For example, the authors of \cite{Hamilton2019} indicated that mixing individuals at different levels of organizational hierarchy should generally be avoided, as power differentials could hamper open discussion. However, our experience with the conducted focus groups was that both directors and IACS engineers demonstrated a strong understanding of cybersecurity challenges in IACSs, and actively contributed to the discussion. Nonetheless, it may be beneficial in future research to hold focus group discussions with participants from the same organizational level to further encourage open dialogue.  

It would also be valuable to include IACS owners in the focus group discussions, particularly given the findings reported in \cite{Norway2024}, where the authors conducted a series of interviews with IACS owners to identify the challenges and issues they face regarding cybersecurity. The results they obtained are in some parts similar to ours, with the key difference that IACS system integrators tend to place greater responsibility for poor cybersecurity on IACS owners, rather than IACS equipment suppliers.

\section{Conclusion and Future Work}
\label{sec:sec6}
This paper examined how Croatian IACS system integrators perceive cybersecurity requirements, implementation practices, incident response responsibilities, and regulatory pressure in industrial environments. The focus group participants report that most organizations adopt a reactive approach to cybersecurity incidents, with unclear responsibilities for managing incidents, often resulting from poorly defined contracts. 

Second, most cybersecurity measures implemented in IACSs are insufficient; while IT cybersecurity equipment is occasionally installed, the deployment of IACS-specific security solutions remains rare. Furthermore, there is a widespread lack of cybersecurity awareness across all IACS sectors, although larger companies show a slightly higher level of preparedness. A difference in cybersecurity awareness between domestic and foreign companies was also observed, with foreign companies generally more willing to invest in cybersecurity measures. What is also interesting, some participants reported that IACS owners often conceal cybersecurity incidents to avoid potential consequences. Finally, opinions on legislation are divided: some participants believe that regulatory frameworks will improve cybersecurity in IACSs, while others believe that most companies will try to meet only the minimum compliance requirements.

The results of this study can serve as a basis for developing more usable cybersecurity practices and for better understanding of key challenges that IACS system integrators and owners face when dealing with cybersecurity. In the future, we will investigate technical challenges faced by IACS system integrators when implementing both IACS-specific and IT cybersecurity solutions through additional focus groups. The technical challenges include issues with legacy systems, patch management in IACSs, and compatibility of cybersecurity measures from different vendors.  We also plan to analyze challenges related to employee training, financial challenges, and differences between IT and IACS cybersecurity practices. 


  \bibliographystyle{elsarticle-num} 
  \bibliography{bibliography_file}



\appendix
\section{Participant quotations}
\setcounter{table}{0}
\begin{table*}
\small
\caption{Quotations regarding cybersecurity awareness}
\label{tab:cybersecurity_awareness}
\centering
\resizebox{\textwidth}{!}{
\begin{tabular}{p{1.6cm} p{15.42cm}}
\hline
\textbf{Subtheme} & \textbf{Quote} \\
\hline
\multirow{4}{=}{\textbf{Non existent /low cybersecurity awareness}}  & (Q1)  "Honestly, most IACS owners do not have a clear understanding of what cybersecurity actually means." \\
& (Q2) "Awareness of cybersecurity risks and security remains quite low, especially among management." \\
& (Q3) "Few people in the IACS environment are even aware of or understand the importance of cybersecurity. That's another big challenge." \\
& (Q4) "Companies are doing well in areas like functional safety and HAZOP, where they have regular training and projects, but when it comes to cybersecurity, awareness is significantly lower." \\
& (Q5) "Besides that, they think they are not a target because there are more attractive targets. For example, when it comes to gas cogeneration plants , they believe the probability of attacks is low. Who would attack us when there are many more interesting targets? You can't extort us when we have no money." \\
& (Q6) "Right now, everyone here is just slowly worrying, keeping an eye on things, and watching to see who will implement something first." \\
& (Q7) "For new IACSs, it is easier because cybersecurity solutions can be planned in advance and included in the budget. However, this is still not commonly done, and so far, we have not had such requirements in contracts."  \\
\hline
\multirow{3}{=}{\textbf{IACS small vs. large companies}} & (Q8) "It's not like that everywhere. There are industries where awareness of risks is high, so the measures are also at a high level. An example is the pharmaceutical industry and parts of the energy sector, while, on the other hand, water and wastewater systems are a standard example of the opposite."  \\
& (Q9) "The problem is further worsened by the lack of trained technical teams, especially among smaller clients. Large organizations typically have engineers or teams specialized in automation, which makes the implementation of security measures easier. Smaller systems, on the other hand, usually lack such support." \\
& (Q10) "In companies that have specialized cybersecurity teams, discussions about technical requirements and changes are made easier." \\
& (Q11) "Some of the larger clients have had cybersecurity programs in place for four or five years, but only now are they beginning to set concrete requirements for their suppliers." \\
& (Q12) "Most of their time and a great deal of money have been devoted to this effort, but only now are concrete requirements regarding personnel, systems, and solutions emerging. Progress is very slow, and some have not even begun implementation yet." \\
\hline
\multirow{3}{=}{\textbf{IACS domestic vs. foreign companies}} & (Q13) "Some foreign companies that implemented systems in (domestically), such as those from (regions with higher cybersecurity awareness), were somewhat more aware of cybersecurity and incorporated certain security measures. " \\
& (Q14) "Regarding remote access by foreign companies, the process is quite controlled, the screen is recorded, and access permissions are granted for a few hours. This ensures monitoring of all activities, but at the same time, it can slow down work in urgent situations." \\
& (Q15) "Especially in regions like (\textit{developing countries}), we encounter situations where people do not understand even the basic concepts of cybersecurity." \\
\hline
\end{tabular}
}
\end{table*}

\begin{table*}
\small

\caption{Quotations regarding response to threats}
\label{tab:cybersecurity_Response_to_Threats}
\centering
\resizebox{\textwidth}{!}{
\begin{tabular}{p{1.6cm} p{15.42cm}}
\hline
\textbf{Subtheme} & \textbf{Quote} \\
\hline
\multirow{2}{=}{\textbf{Reactive response}}  & (Q16) "At the end of the day, the key thing is business continuity; when something happens, there is no focus on cost or responsibility, what matters is getting the system back up and running as quickly as possible." \\
& (Q17) "So, in the case of an attack, a root cause analysis is not immediately conducted. Instead, we most likely go straight to resolve the issue and put out the fire as quickly as possible. If we encounter problems we cannot solve, we escalate them to services and external companies." \\
& (Q18) "If a problem occurs, it is expected that we go on-site and resolve the situation by restoring backups, installing a new system image, and similar actions." \\
\hline
\multirow{2}{=}{\textbf{Responsi\-bility issues}} & (Q19) "Once the client takes over the system, the responsibility for its monitoring and response to threats falls on them."\\
& (Q20) "Currently, we provide advisory reports through our company's website to inform clients about potential threats.
If a client has contracts such as a service level agreement (SLA), we notify them of any issues. If no such agreement exists, the client is responsible for monitoring the situation. We have not yet fully defined our SLA service; our maintenance contracts are primarily related to on-site equipment." \\
& (Q21)  "It is crucial to understand that security management must be the owner's responsibility, not the integrator's task." \\
& (Q22) "There have been cases where an incident was detected, but most often it was due to devices brought in by the customer, such as computers with already infected files."
 \\
& (Q23) "SOCs are usually a service provided by an external company, which comes with its own challenges. They send alerts if something happens, but due to the high number of false positives and negatives, it is difficult to identify what actually occurred and respond appropriately." \\
\hline
\multirow{2}{=}{\textbf{Covering up incidents}}& (Q24) "Many still cover up incidents. Over 70-80\% of attacks occur because previous incidents were concealed, which only worsens the problem." \\
& (Q25) "One ransomware incident on a workstation led the owner to attempt justifying the costs of replacing part of the system as the installation of 100 meters of cable." \\
\hline
\end{tabular}
}
\end{table*}

\begin{table*}
\small

\caption{Quotations regarding technical measures in IACSs}
\label{tab:cybersecurity_Technical_Measures}
\centering
\resizebox{\textwidth}{!}{

\begin{tabular}{p{1.6cm} p{15.42cm}}
\hline
\textbf{Subtheme} & \textbf{Quote} \\
\hline
\multirow{2}{=}{\textbf{No measures}}  & (Q26) "Unfortunately, in most companies, there are almost no technical measures in place." \\
& (Q27) "We generally do not maintain a balance between requirements, as availability is prioritized highly, while cybersecurity is given very little importance. In fact, cybersecurity is not considered as requirement at all." \\
& (Q28) "Some companies, particularly larger ones, attempt to implement certain measures on their own initiative. However, most of the IACSs we visit have no cybersecurity measures in place at all." \\
\hline
\textbf{IT measures} & (Q29) "Now, for example, updates for Windows systems are almost automatic, but that’s with newer systems. […]  but generally, the basic issue is that we need to start with the basics. The best firewall means nothing if it's not correctly configured and if it has an "admin admin" password."\\
& (Q30) "We don’t have much experience with direct cyberattacks, but the most commonly used solution in practice are intrusion detection systems (IDSs) at the IT level, and honestly, we are happy that this is not in our domain." \\
& (Q31) "Recently, we had a meeting with a medium-sized client. Their requirements are quite basic, like minimal hardening, logging, blocking changes and access to certain records, and separating passwords for different classes of users [..] Generally, clients usually start with smaller requirements, and only later do they expand them." 
\\
\hline
\multirow{2}{=}{\textbf{IACS-specific measures}}  & (Q32) "We mainly install firewalls, VPNs, and VLANs." \\
& (Q33) "For example, the improvement of remote access started a bit earlier because previously it was handled quite unorganized […]. Similarly, about two years ago, they started using ******* solutions for remote access in the industry. ******* provides both hardware and software for this purpose." \\
&( Q34) "Today, some organizations have separate Active Directory systems, one for business (IT) and another for processes (IACS). This division allows for specific security policies tailored to each sector." \\
\hline
\end{tabular}
}
\end{table*}

\begin{table*}
\small

\caption{Quotations regarding legislation impact}
\label{tab:cybersecurity_Legislation}
\centering
\resizebox{\textwidth}{!}{
\begin{tabular}{p{1.6cm} p{15.42cm}}
\hline
\textbf{Subtheme} & \textbf{Quote} \\
\hline
\multirow{2}{=}{\textbf{Minimal compliance with legislative requirements}}  & (Q35) "A law on cybersecurity was already introduced in some form back in 2018, but it did not compel companies to make significant changes. In the end, it is very likely that people will aim only to meet the bare legal minimum. They will try to obtain the necessary paperwork and nothing more." \\
& (Q36) "The approach is quite superficial; everyone mainly wants to handle the paperwork and be compliant with laws and requirements, and that's it, just the basic minimum. Similarly, cybersecurity requirements rarely appear in tenders, so the implementation of security measures is often a secondary concern." \\
& (Q37) "I'm afraid that the entire regulation will end up being handled by lawyers, not security experts." \\
& (Q38) "A similar situation once occurred with Safety HAZOP analyses. Some investors kept looking for companies that would declare them compliant, until one company finally did, while the other three companies that had conducted the analyses kept identifying errors and non-compliances. This kind of situation should be avoided through standardization and adherence to established norms." \\
\hline
\multirow{4}{=}{\textbf{Legisla\-tion driving cybersecurity efforts}} & (Q39) "At this moment, there are very few requirements for cybersecurity, and in general, regulation will most likely force companies to address it." \\
& (Q40) "Absolutely. As soon as they are compelled by law and there is a risk of penalties, it leads to increased interest in implementing security measures." \\
& (Q41) "Awareness of the importance of security support is growing, and new regulations, such as NIS2 are increasingly encouraging manufacturers to listen to what customers are asking for and need." \\
& (Q42) "The assessment must be tied to something concrete, but it is expected that some will try to exploit the chaos in the market due to time limitations. However, we expect that serious users will choose those who work in accordance with the law and standards." \\
\hline
\end{tabular}
}
\end{table*}


\end{document}